%% file: M87UCDs.tex
\documentclass[fleqn,usenatbib]{mnras}

\usepackage{booktabs}

\usepackage{newtxtext,newtxmath}

\usepackage[T1]{fontenc}
\usepackage{ae,aecompl}
\usepackage{xcolor}

\usepackage{graphicx}	
\usepackage{amsmath}	
\usepackage{amssymb}	
\usepackage{threeparttable}

\input{extradefs}

\title[M87 compact objects]{Keck Cosmic Web Imager (KCWI) Spectra of Globular Clusters and  Ultra Compact Dwarfs in the Halo of M87}

\author[D. A. Forbes et al.]{Duncan A. Forbes,$^{1}$\thanks{E-mail: dforbes@swin.edu.au}
Anna Ferr\'e-Mateu$^{1,2}$, Mark Durr\'{e}$^{1}$, Jean P. Brodie$^{3}$ and
\newauthor
Aaron J. Romanowsky$^{3,4}$
\\
$^{1}$Centre for Astrophysics \& Supercomputing, Swinburne University, Hawthorn VIC 3122, Australia\\
$^{2}$Institut de Ciencies del Cosmos (ICCUB), Universitat de Barcelona (IEEC-UB), E02028 Barcelona, Spain\\
$^{3}$University of California Observatories, 1156 High St., Santa Cruz, CA 95064, USA\\
$^{4}$Department of Physics \& Astronomy, San Jos\'e State University, San Jose, CA 95192, USA
}

\date{Accepted XXX. Received YYY; in original form ZZZ}

\pubyear{2020}

\begin{document}
\label{firstpage}
\pagerange{\pageref{firstpage}--\pageref{lastpage}}
\maketitle

\begin{abstract}

Using the Keck Cosmic Web Imager we obtain spectra of several globular clusters (GCs), ultra compact dwarfs (UCDs) and the inner halo starlight of M87, at a similar projected galactocentric  radius of $\sim$5 kpc. This enables us, for the first time, to apply the same stellar population analysis to the GCs, UCDs and starlight consistently to derive ages, metallicities and alpha-element abundances in M87. 
We find evidence for a dual stellar population in the M87 halo light, i.e an $\sim$80\% component by mass which is old and metal-rich 
and a $\sim$20\% component which is old but metal-poor.
Two red GCs share similar stellar populations to the halo light 
suggesting they may have formed contemporaneously with the dominant halo component.
Three UCDs, and one blue GC, have similar stellar populations, with younger mean ages, lower metallicities and near solar 
alpha-element abundances. 
Combined with literature data, our findings are consistent with the scenario that UCDs are the remnant nucleus of a stripped galaxy. 
We further investigate the discrepancy in the literature for M87's kinematics at large radii, favouring a declining velocity dispersion profile. This work has highlighted the need for more self-consistent studies of galaxy halos.


\end{abstract}





\section{Introduction}

The stellar halos of early-type galaxies provide important clues to their assembly history. 
As dynamical times are longer in the outer halos, they better preserve the signatures of past mergers and accretion. Simulations predict that more massive galaxies have an increasing fraction of stars that have been accreted (\citealt{2010ApJ...725.2312O}; \citealt{2013MNRAS.434.3348C}; \citealt{2014MNRAS.444..237P}), reaching around 90\% for the most massive galaxies. Accretion of material at more recent times is likely to be in the form of minor mergers. Such mergers tend to deposit stars, and globular clusters, into the outer regions of the host galaxy \citep{2019MNRAS.487..318K}. 
These accreted stars will tend to be old, metal-poor and alpha-element  enriched  (\citealt{2005ApJ...635..931B}; \citealt{2019MNRAS.482.3426M}). 
Today's early-type galaxies will, in general, be a mix of both in-situ and ex-situ formed stars and globular clusters \citep{2018MNRAS.479.4760F}. 

As well as field stars and globular clusters (GCs), stellar halos may host ultra compact dwarfs (UCDs). Ultra compact dwarfs were first identified roughly twenty years ago \citep{1999A&AS..138...55H,2000PASA...17..227D}
in the halo of NGC~1399 and the surrounding Fornax cluster. In visual appearance they resemble large, luminous GCs but the defining characteristics of a UCD vary in the literature. Here we adopt the criteria of an effective radius of R$_e$ $>$ 10 pc combined with M$_V$ $<$ --9 (which corresponds to a stellar mass of $\ge$10$^6$ M$_{\odot}$). With these criteria, the Milky Way GC NGC~2419 (with R$_e$ = 21 pc and M$_V$ = --9.4) would be regarded as a UCD ($\omega$ Cen meets the luminosity criterion with M$_V$ = --10.3 but not the size one with R$_e$ = 7.6 pc). 


Globular cluster systems generally reveal two subpopulations when examined in terms of their colours or metallicities \citep{2006ARA&A..44..193B}. The red, or metal-rich, subpopulation is thought to be associated with the field stars of early-type galaxies since they have similar mean colours \citep{2001MNRAS.322..257F},  kinematics \citep{2013MNRAS.428..389P} and surface density profiles 
\citep{2012MNRAS.425...66F}. The blue, or metal-poor, GCs may follow the radial distribution of the halo as traced by the hot gas \citep{2012MNRAS.425...66F} and they reveal kinematics that are less correlated with the host galaxy stars \citep{2013MNRAS.428..389P}. 


The stellar halos of a few nearby (D $\le$ 10 Mpc) galaxies have been resolved into individual stars with HST imaging. The current status is summarised by 
\citet{2020arXiv200101670C} in their table 3 which lists 10 early-type and 5 late-type galaxies. A diversity of halo properties are seen. For example, the halo of the Sombrero galaxy, probed out to 17R$_e$, is dominated by metal-rich stars with only a negligible contribution of stars with metallicity [Z/H] $<$ --1. This suggests a single major merger in its assembly history. In contrast, both NGC 3379 and NGC 5128 reveal a transition to a metal-poor halo at radii of around 10--15R$_e$. 
In the case of NGC 3115 the metal-poor stars have a mean metallicity approaching that of its metal-poor GC subpopulation with \cite{Peacock2015} concluding that {\it ``This is the strongest evidence to date that globular clusters trace the stellar populations in the halos of early-type galaxies."}. They further suggested that GCs can be used as {\it ``chemo-dynamical tracers of the stellar halos of more distant galaxies."} 

We also note the work of  \citet{2010A&A...524A..71B} (not included in table 3 of
\citealt{2020arXiv200101670C}) 
that obtained deep HST imaging of individual red giants in the inner halo of M87, finding a predominately metal-rich halo as expected but also a metal-poor tail in the metallicity distribution of stars down to [Z/H] $\sim$ --2. At the distance of the Virgo cluster (D = 16.5 Mpc) such observations become increasingly difficult requiring long exposure times with the HST.  
For more distant galaxies, one must use the integrated properties of stellar halos and this is now being carried out with integral field units on large telescopes (e.g. \citealt{2019ApJ...878..129F}).


One challenge in comparing the stellar populations of the integrated starlight in early-type galaxy halos and associated compact objects is that usually they are studied with different instruments and using different stellar population models. Systematic differences between models can be particularly acute at the highest metallicities (see for example figure 5 of \citealt{2011ApJ...729..129M}), which is the regime of metal-rich GCs and the field stars of giant early-type galaxies. Recently, integral field units that are designed for low surface brightness targets have been installed on 8--10m telescopes, i.e. MUSE \citep{2010SPIE.7735E..08B}
on the VLT and KCWI \citep{Martin2010c} on Keck II. Such instruments can be used to probe both the low surface brightness outer regions of galaxy halos while capturing several halo compact objects in the same field-of-view. We note that SLUGGS  \citep{2014ApJ...796...52B} was the first survey to obtain large numbers of GC spectra and their host galaxy starlight spectra in the same observation using the DEIMOS multi-slit spectrograph on Keck II as a `pseudo integral field unit'. However, the stellar population properties of GCs and field stars were derived using different methods, e.g. \citet{2015MNRAS.451.2625P}. 
In the ongoing Fornax3D survey \citep{2018A&A...616A.121S} the study of both stellar halos of early-type galaxies in the Fornax cluster, and their systems of GCs, can be obtained from the same MUSE pointing along with self-consistent stellar population modelling.

Here we study the inner halo of M87 (the central cD galaxy of the Virgo cluster) using the integral field unit KCWI on the Keck II telescope. 
As well as field stars, our single KCWI pointing includes several GCs and UCDs. Thus we are able to obtain spectra of the integrated galaxy halo and these compact objects in the same exposure with the same instrument, thereby avoiding any relative calibration issues when studying both halo field stars and compact objects. Our study is also the first to compare their stellar populations using the same methodology.  In this work we adopt a distance to M87 of 16.5 Mpc (m--M = 31.1) where 10 arcsec = 801 pc.

\section{Observations and Data Reduction}

\subsection{\textit{HST/ACS} Imaging}

\textit{HST} images of the M87 halo were obtained from program ID 10543. 
These used the ACS/WFC instrument, with 49 individual images each in the F606W and F814W filters. The images have a spatial resolution of $0.05\arcsec$. The FITS drizzled files were registered and combined using the SWARP\footnote{\url{http://astromatic.net/software/swarp/}} software. 

\begin{figure}
	\centering
	\includegraphics[width=1\linewidth]{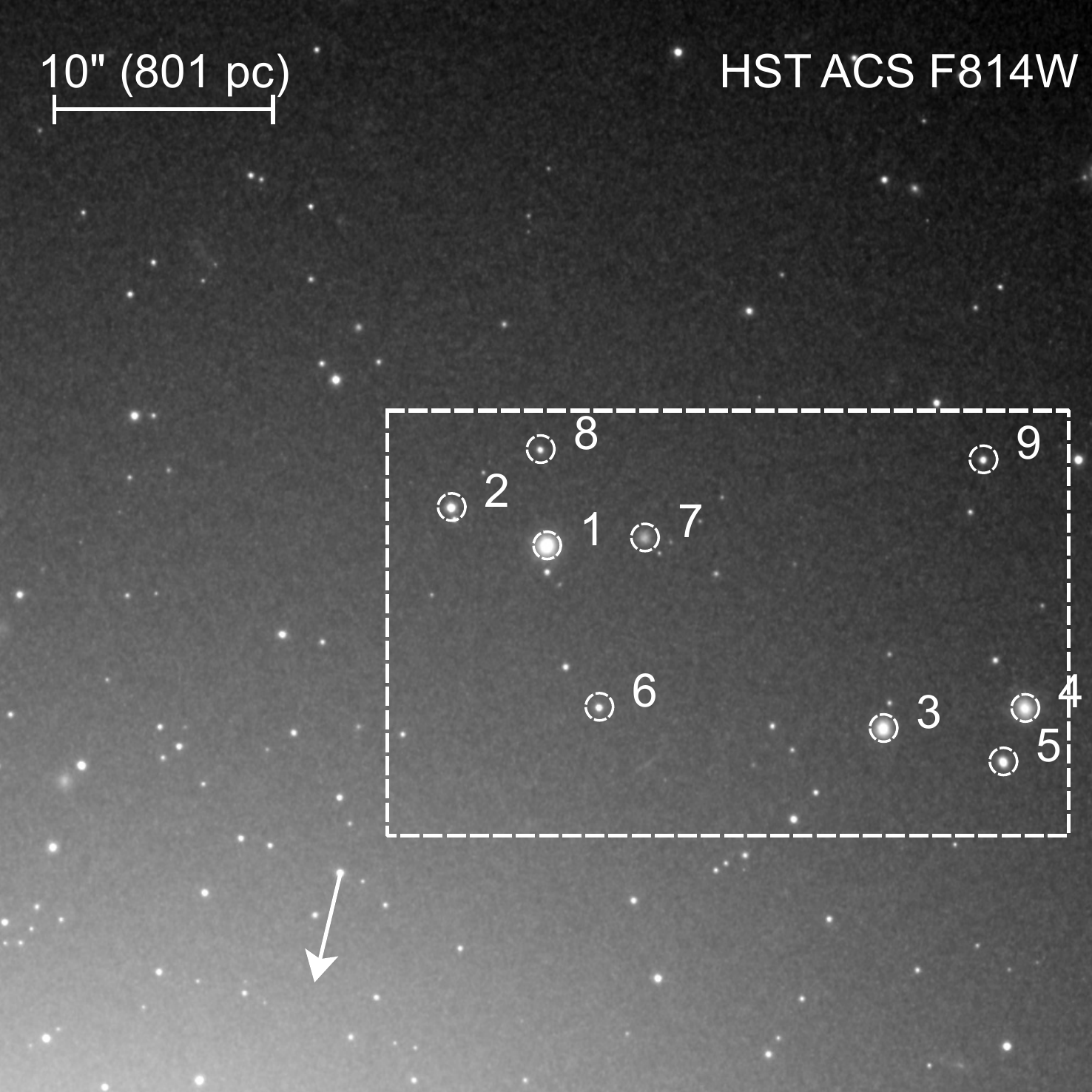}
	\caption{\textit{HST/ACS} F814W pointing to the NW of the M87 nucleus 
	(the arrow indicates the direction of the M87 nucleus).  North is up and East is left. The white box is the FOV of our Keck/KCWI observations with the compact objects indicated.}
	\label{fig:m87cfieldf814w}
\end{figure}

We show in Figure \ref{fig:m87cfieldf814w} the resulting M87 inner halo field located $\sim$5 kpc from the centre of M87. It reveals numerous compact objects. The white dashed box shows the field of view of our Keck/KCWI observations, and the white arrow shows the direction to the centre of M87. The 9 brightest compact objects are numbered as used in the rest of the paper.
The background gradient from bottom left to top right is the stellar halo light of M87. 

\begin{figure}
	\centering
	\includegraphics[width=1\linewidth]{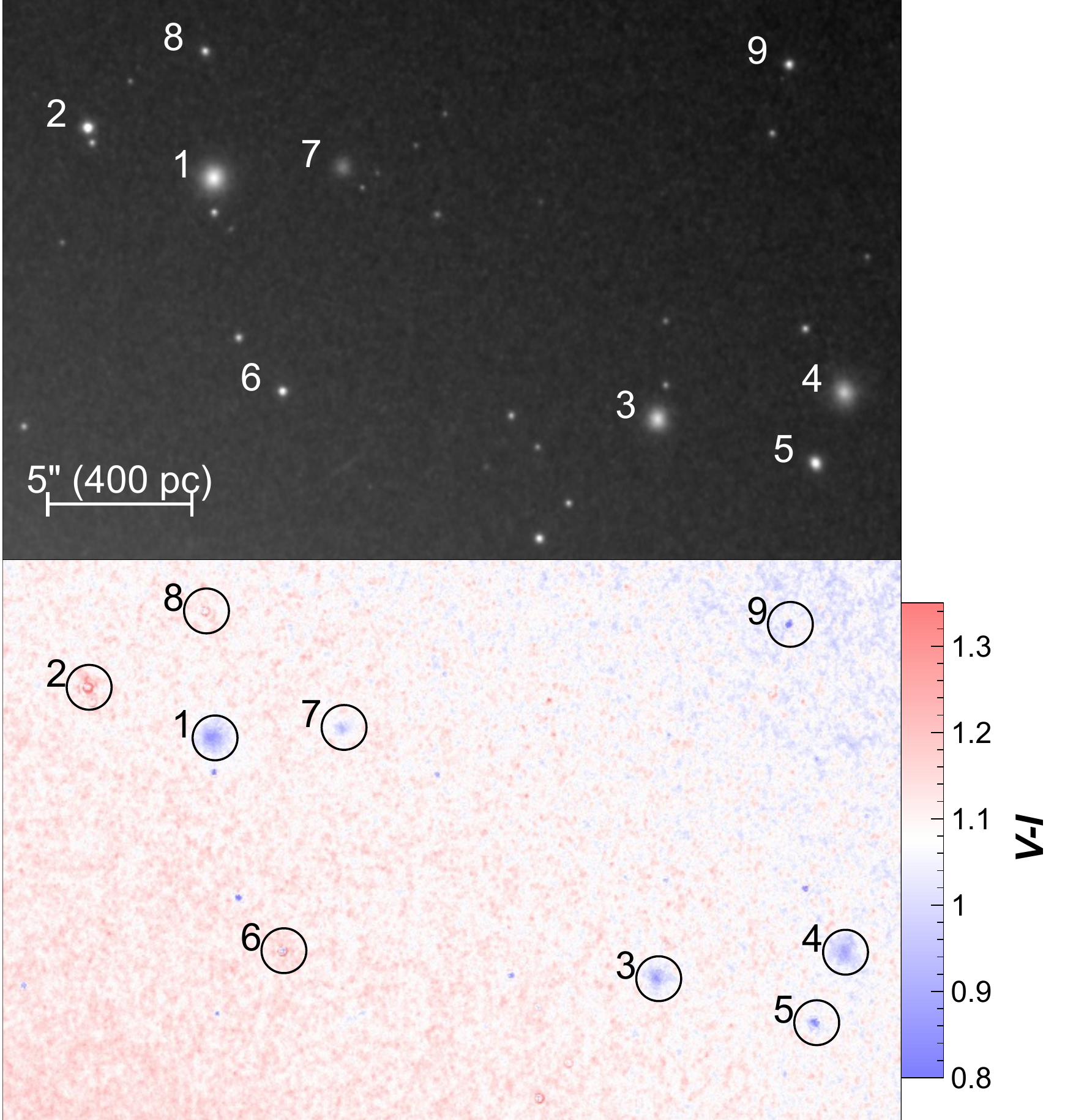}
	\caption{Zoom-in image (corresponding to our Keck/KCWI FOV) with compact objects indicated. {\it Top} The HST/ACS F814W image showing that some objects are clearly extended. {\it Bottom} The HST/ACS \textit{V-I} colour image showing the range of colours for the compact objects.}
	\label{fig:m87cfieldobjects}
\end{figure}

Figure \ref{fig:m87cfieldobjects} shows a zoom-in of the HST/ACS pointing corresponding to our Keck/KCWI field-of-view (FOV) and colour map. The small background colour gradient, of $\sim$0.2 mag, from bottom left to top right is likely due to the radial metallicity gradient in the halo of M87. The compact objects reveal a range of colours, with objects 2, 6 and 8 being red (similar to the colour of M87's inner halo starlight) while the others appear blue. 




We fit each compact object with a \ssc{} profile and measure its \ssc{} index n. The fluxes for each object are converted to relative magnitudes using the \textit{HST} ACS zero-points, as given in the image file header. The corresponding Johnson V and I magnitudes (on the Vega system) are computed using the methods from \citet{Holtzman1995}, with the coefficients from \citet{Sirianni2005}; the resulting V--I colours have values from 0.9 to 1.3, consistent with old stellar populations with a range of metallicities. Table \ref{tbl:M87CObjectData1} gives the object coordinates in J2000 
(the galaxy centre is located at
RA = 187.705930, Dec = 12.391123), along with the corresponding ID from the catalogues of 
\citet{2009ApJS..180...54J} or 
\citet{Brodie2011}. We also list 
our measured \ssc{} index n parameter from the I band image. 
Table \ref{tbl:M87CObjectData2} 
gives our assigned object identification (i.e. GC or UCD) based on its physical size (taken directly from either \citealt{2009ApJS..180...54J} or \citealt{Brodie2011})  
and our measured V band absolute magnitude.
The table also lists their total apparent magnitudes and V--I colour.

\begin{table}
\caption{M87 compact objects: coordinates, IDs and indices.}
\label{tbl:M87CObjectData1}
\begin{tabular}{@{}ccclr}
\toprule
Object & RA         & Dec       & Other & S\'{e}rsic n \\
  ID   &  (J2000)          &     (J2000)      &   ID  & Index        \\ \midrule
1      & 187.698777 & 12.408740 & S928    & 0.83 \\
2      & 187.700020 & 12.409172 &   J1643      & 0.82\\
3      & 187.694393 & 12.406416 & S8006   & 0.78    \\
4      & 187.692549 & 12.406671 & S8005   & 0.88    \\
5      & 187.692499 & 12.405991 &   J1697      & 0.72\\
6      & 187.698099 & 12.406685 &  J1543       & 0.81\\
7      & 187.697502 & 12.408845 & H46484  & 0.63\\
8      & 187.698863 & 12.409964 &  J1729       & 0.97\\
9      & 187.693092 & 12.409835 &   J1896      & 0.73\\
\bottomrule
\end{tabular}
\begin{tablenotes}
\item Other ID from catalogues of 
\citet{2009ApJS..180...54J} or 
\cite{Brodie2011}. 
\end{tablenotes}
\end{table}

\begin{table*}
	\centering
	\caption{M87 compact objects: classifications, sizes, magnitudes and colours.}
	\label{tbl:M87CObjectData2}
	\footnotesize
	\begin{tabular}{@{}ccccrrrrr@{}}
		\toprule
Object & Classification & M$_V$  & R$_e$       & F606W & F814W & V     & I     & V-I   \\
ID     &                & (mag)  & (pc)        & (mag) & (mag) & (mag) & (mag) & (mag) \\ \midrule
1      & UCD            & -11.42 & 36.3       & 19.82 & 20.10 & 19.68 & 18.76 & 0.92  \\
2      & GC             & -10.18 & 1.90 & 20.97 & 20.97 & 20.92 & 19.63 & 1.30  \\
3      & UCD            & -10.74 & 31.7       & 20.49 & 20.74 & 20.36 & 19.41 & 0.95  \\
4      & UCD            & -10.79 & 36.9       & 20.45 & 20.73 & 20.31 & 19.39 & 0.92  \\
5      & GC             & -10.12 & 6.68 & 21.11 & 21.36 & 20.98 & 20.02 & 0.96  \\
6      & GC             & -9.26  & 1.98 & 21.91 & 21.97 & 21.84 & 20.63 & 1.21  \\
7      & UCD            & -9.20  & 39.1       & 22.04 & 22.32 & 21.90 & 20.98 & 0.92  \\
8      & GC             & -8.67  & 2.49 & 22.49 & 22.54 & 22.43 & 21.20 & 1.23  \\
9      & GC             & -9.21  & 3.70 & 22.03 & 22.30 & 21.89 & 20.96 & 0.93 \\ \bottomrule
	\end{tabular}
\end{table*}

\subsection{Keck/KCWI Observations}

A section of the HST/ACS M87 halo pointing  was observed with the Keck Cosmic Web Imager \citep[KCWI;][]{Martin2010c} as part of our Keck Program (ID U250). 
We obtained the observations on 2018 May 9, using the BL grating and large slicer with a central wavelength of 4550~\AA{} and a usable wavelength range of 3600--5700 \AA. This  produced a spectral resolution R$\sim$900 (5.06~\AA{}/pix at the central wavelength), with a field of view of $33\arcsec \times 20.4\arcsec$ and a spatial resolution of $1.35\arcsec \times 0.29\arcsec$. Calibration frames (arcs, flats, geometric bars) were also observed as well as observations of the reference star Feige 67 with the same setup. The data were reduced using the KCWI pipeline
which performs a full data reduction, delivering flux-calibrated datacubes. 

\begin{figure}
	\centering
	\includegraphics[width=1.0\linewidth]{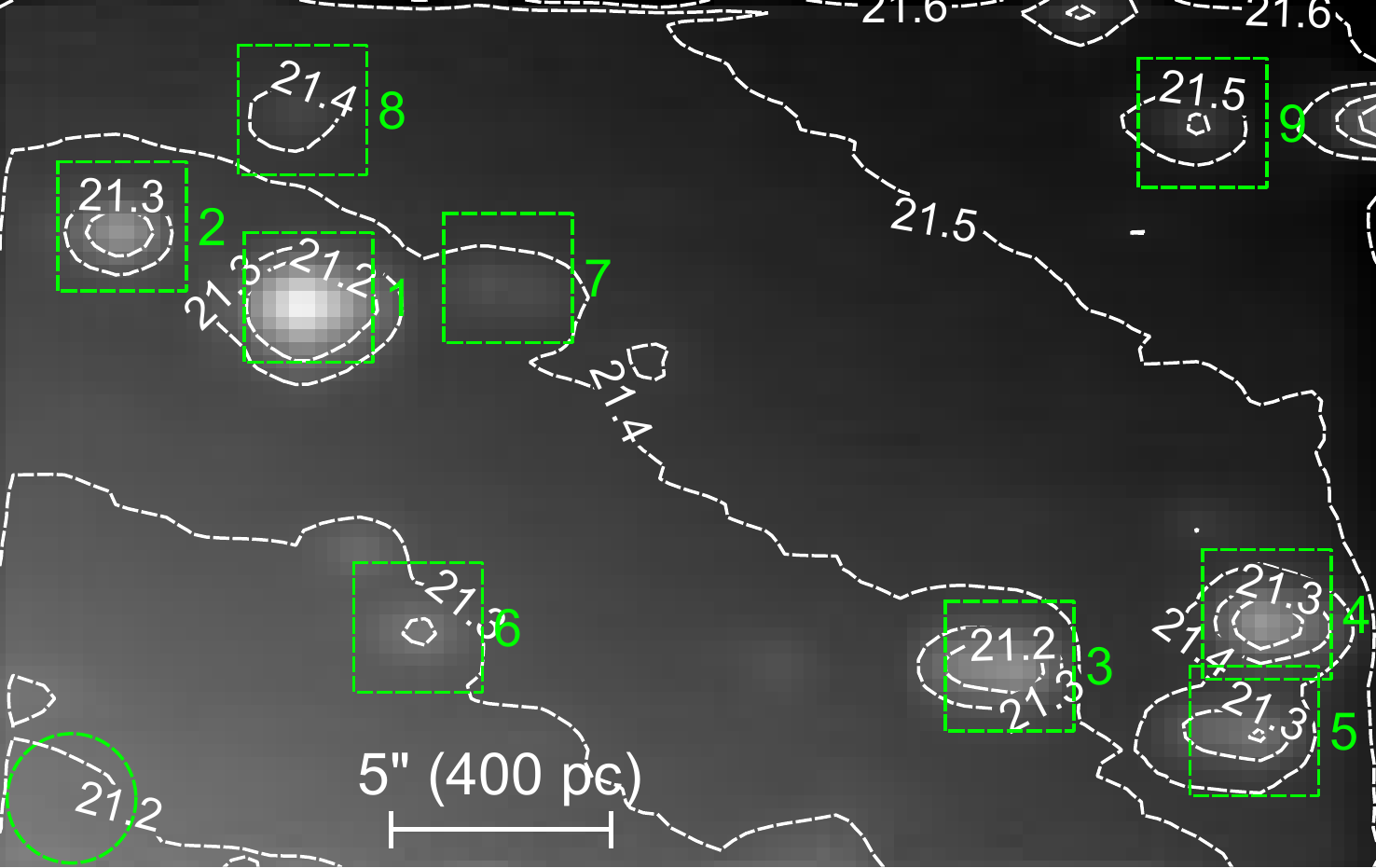}
	\caption{Keck/KCWI image centered around 4380\AA ~(corresponding to the B band). The compact objects are numbered and observed B band surface brightness contours in  magnitudes/arcsec$^2$ are indicated. The visible elongation of the objects is an artifact of the rebinning process. The green dashed circle, in the lower left,  indicates the aperture used to extract the galaxy halo light.}
	\label{fig:m87ckcwifield}
\end{figure}

Seven frames were obtained (5 of 600 sec and 2 of 300 sec) at a short-axis position angle of 0$\degr$. The resulting cubes have $28 \times 96 \times 2569$ pixels. The cubes are re-binned for convenience so that each pixel is square (i.e. $ 0.29\arcsec $ on a  side), using a bespoke code. This produced a spatial field of view of $131 \times 96$ pixels. We combined the five 600 sec exposures together  (the 300 sec exposures have too poor signal to noise). The observation tracking was good enough so that no re-registration was required. The resulting cube was trimmed to 3560-5560 \AA{} to remove spectral vignetting on the CCD image, and to remove the 5577 \AA{} skyline. Figure \ref{fig:m87ckcwifield} shows a KCWI image, with the same FOV as 
Fig. \ref{fig:m87cfieldobjects},  
centered
around a wavelength of 4380~\AA{} (corresponding roughly to a Johnson B band filter). Overplotted are the GCs and UCDs, and the location of the extracted M87 halo light.


Using \texttt{QFitsView}, we extracted the spectrum of each object from the data cube within an aperture of 1 R$_e$ (as measured with a \ssc{} fit from the image created by collapsing the cube along the wavelength axis). This aperture was found to be a good compromise between signal and added noise. The background, consisting of galaxy halo and sky light, was subtracted using an annulus around each object (with light from other objects masked out). We also extracted a galaxy halo light spectrum with an aperture of 5 pixels radius centered at RA = 187.700296, Dec = 12.405650 (J2000) (some 56 arcsec or 4.5 kpc from the galaxy centre), as indicated on Fig. \ref{fig:m87ckcwifield} by a green dashed circle. The background for this spectrum was taken from an aperture of the same size at the top right of the cube. 
We investigated fitting a smooth 2D surface to the image and using it for  background subtraction but this gave inferior results compared to the simple same-sized aperture approach. 

In this work we assume an effective radius (R$_e$) for M87 of 87 arcsec as measured by \citet{2017MNRAS.464.4611F} from a single Sersic fit to 3.6$\mu$m imaging. Thus our pointing lies at a projected radius of $\sim$0.6~R$_e$. We also note the large range in R$_e$ values for M87 in the literature (see discussion in \citealt{2017MNRAS.464.4611F}).


\section{Stellar Population Analysis} 
\subsection{Methodology}

We use the MILES single stellar population (SSP) models of
with BaSTI isochrones \citep{2004ApJ...612..168P} and a Kroupa Universal IMF to obtain our stellar population parameters (\citealt{2010MNRAS.404.1639V}; \citealt{2015MNRAS.449.1177V}). The templates cover a range of metallicites from [Z/H] = $-$2.27 to $+$0.40\, dex and ages from 0.03 to 14\,Gyr. These models also allow for two different alpha-element abundances, i.e. scaled-solar or super-solar ([$\alpha$/Fe]$=+$0.4\,dex). We adopt a two-step approach. First, we obtain approximate alpha-element abundance  values using the classical approach of measuring Magnesium and Iron absorption line indices for each galaxy and display them in an index-index model grid. In particular, we measure Mg$_b$ and the combination of the Fe5270 and Fe5335 lines, $<$Fe$>$.
The line indices are measured with LIS5
\citep{2010MNRAS.404.1639V}, which matches the spectral resolution of our data. 

\begin{figure}
	\centering
	\includegraphics[width=1\linewidth]{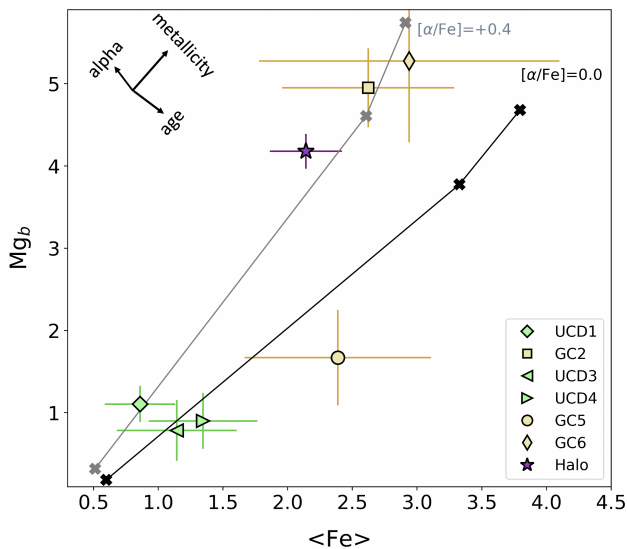}
	\caption{Mg$_b$ and $<$Fe$>$ = (Fe5270+Fe5335)/2 line indices on stellar population model tracks. Two tracks shown are of age 10 Gyr for two alpha-element abundances, i.e. scaled-solar ([$\alpha$/Fe]=0.0\,dex, black) and super-solar ([$\alpha$/Fe]$=+$0.4\,dex, grey). 
	Metallicities of $-$2.27, 0.06 and 0.4\,dex are indicated. The arrows in the upper left indicate the directions of increasing alpha-element abundances, higher metallicities and younger ages. 
	Our measurements of the halo light of M87, along with those for GCs and UCDs, are overplotted. They are consistent with either scaled-solar or super-solar alpha-element abundances. 
}
	\label{fig:final_grid}
\end{figure}

Fig.~\ref{fig:final_grid}
shows our line indices measurements compared to two SSP tracks with fixed age of 10 Gyr and two different [$\alpha$/Fe] abundance values (i.e. solar and super-solar). The assumption of 10 Gyr is a reasonable one given the range of ages we derive below. We obtain approximate 
alpha-element values (to an accuracy of 0.05 dex) by comparing our measurements with the two tracks. Most of our measurements are fully consistent with one of the two tracks, and for GC6 we perform a simple linear interpolation between the two tracks to estimate its [$\alpha$/Fe] value. 
For GC5 and the halo light, which lie slightly outside of the two tracks to lower and higher alpha-element abundances respectively, we perform a simple extrapolation assuming a linear continuation of the SSPs. This is clearly a somewhat qualitative approach but is sufficient to discriminate between super-solar and solar alpha-element abundance objects. 
From this process we infer that the M87 inner halo light, UCD1, GC2 and GC6 are enhanced, while UCD3, UCD4 and UCD5 are not. 
To estimate the total uncertainty on [$\alpha$/Fe] for each object we combine the errors associated with each method we apply, i.e. pairs of indices (described above) and full spectral fitting (described below).

\begin{figure*}
	\centering
	\includegraphics[width=1\linewidth]{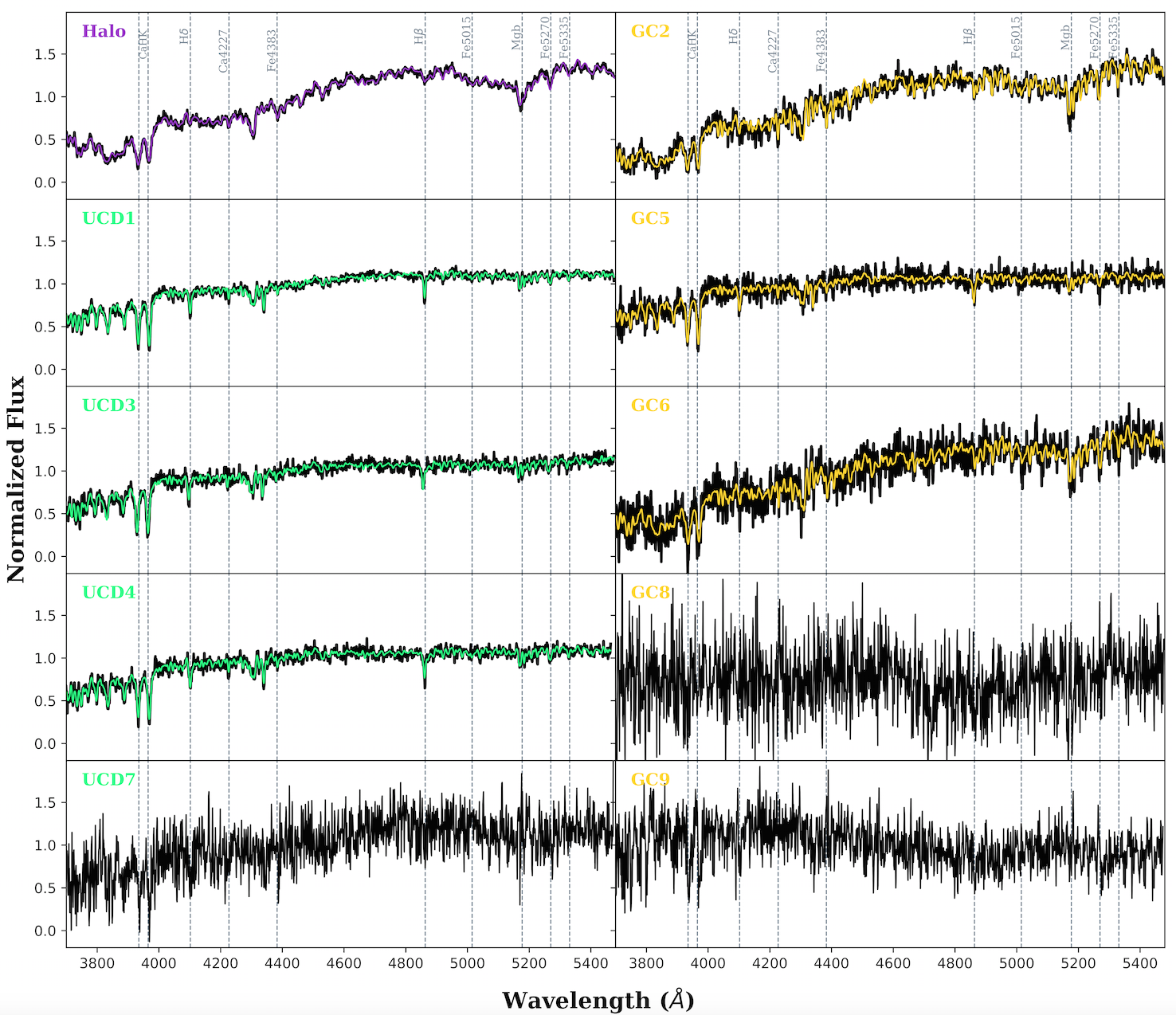}
	\caption{KCWI spectra of the M87 halo starlight and compact objects. The original spectra are shown in black, while the coloured lines show the best fit stellar population models when available. The location of some key absorption lines are indicated.
	}
	\label{fig:spectra}
\end{figure*}

In the second step, we apply the full-spectral-fitting code {\tt pPXF} (Penalized Pixel Fitting; \citealt{2004PASP..116..138C}) with an identical set of SSP models with either scaled-solar or alpha-enhanced templates according to the result from the first step. Figure \ref{fig:spectra} shows the spectra of the M87 halo light and each compact object. For those spectra with sufficient S/N (i.e. objects 1--6), we also show our best fit model from our {\tt pPXF} fitting.

To check the robustness of our results, we also employed another full-spectral-fitting code, {\tt STECKMAP} \citep{2011arXiv1108.4631O}, and obtained metallicities from the classical absorption line index method. Our results for the three methods ({\tt pPXF, STECKMAP} and line indices) are consistent within the uncertainties. We also tested both non-regularized (which is similar to the classical line index approach) and regularized solutions (similar to {\tt STECKMAP}) within {\tt pPXF}. The regularization is obtained as described in the code manual \citep{2017MNRAS.466..798C}. It provides a trade-off between a smoother star formation history and a good fit to the data. Another test we carried out was to only fit each spectrum for wavelengths shorter than 5100~\AA ~(to avoid strong sky emission). This gave consistent results as using the full wavelength coverage. The final quoted uncertainties on all stellar population parameters are the mean of all these approaches combined. 

\begin{table*}
	\centering
	\caption{M87 compact objects: radial velocities, stellar populations and masses. }
	\label{tbl:M87CObjectData3}
	\footnotesize
	\begin{tabular}{@{}ccccrccc@{}}
		\toprule
Object & Identification & S/N & Vel. & Age & [Z/H] & [$\alpha$/Fe] & $\mathrm{M_{*}}$\\
ID     &                & & (km/s) & (Gyr) & (dex) & (dex)   & ($\times$10$^{6}$M$_{\odot}$) \\  \midrule
1      & UCD            & 43 & 1311$\pm$5 &  7.70 $\pm$0.6 & -0.95$\pm$0.12 &  0.40$\pm$0.15 &  5.99$\pm$0.49 \\
2      & GC             & 18 & 1308$\pm$4 & 11.60 $\pm$1.4 &  0.20$\pm$0.12 &  0.40$\pm$0.20 &  4.37$\pm$0.38 \\
3      & UCD            & 24 & 1041$\pm$3 &  9.15 $\pm$1.1 & -0.86$\pm$0.13 &  0.00$\pm$0.30 &  3.19$\pm$0.38 \\
4      & UCD            & 26 & 1910$\pm$6 &  9.36 $\pm$0.6 & -0.79$\pm$0.12 &  0.00$\pm$0.35 &  3.98$\pm$0.54 \\
5      & GC             & 15 &  839$\pm$8 &  8.19 $\pm$0.9 & -0.67$\pm$0.11 & -0.10$\pm$0.30 &  1.90$\pm$0.42 \\
6      & GC             &  8 & 1371$\pm$5 & 10.26 $\pm$1.3 & -0.10$\pm$0.20 &  0.35$\pm$0.40 &  1.29$\pm$0.38 \\
7      & UCD            &  4 & 1571$\pm$5 & --             & --             & --             & -- \\
8      & GC             & -- & --         & --             & --             & --             & -- \\
9      & GC             &  4 & 1630$\pm$9 & --             & --             & --             & -- \\
\hline                         
M87      & Halo         & 34 & 1246$\pm$3 &  12.10 $\pm$1.1 &   0.21$\pm$0.05 & 0.45$\pm$0.10 & -- \\
         & $\sim$80\%   & --   &      --      &  12.00 $\pm$0.5 &   0.16$\pm$0.12 &   --            & --  \\
         & $\sim$20\%   &  --  &    --        &  10.50 $\pm$0.5 &  -0.66$\pm$0.30 &    --           & --  \\
\bottomrule
	\end{tabular}
\end{table*}

\begin{figure}
	\centering
	\includegraphics[width=1\linewidth]{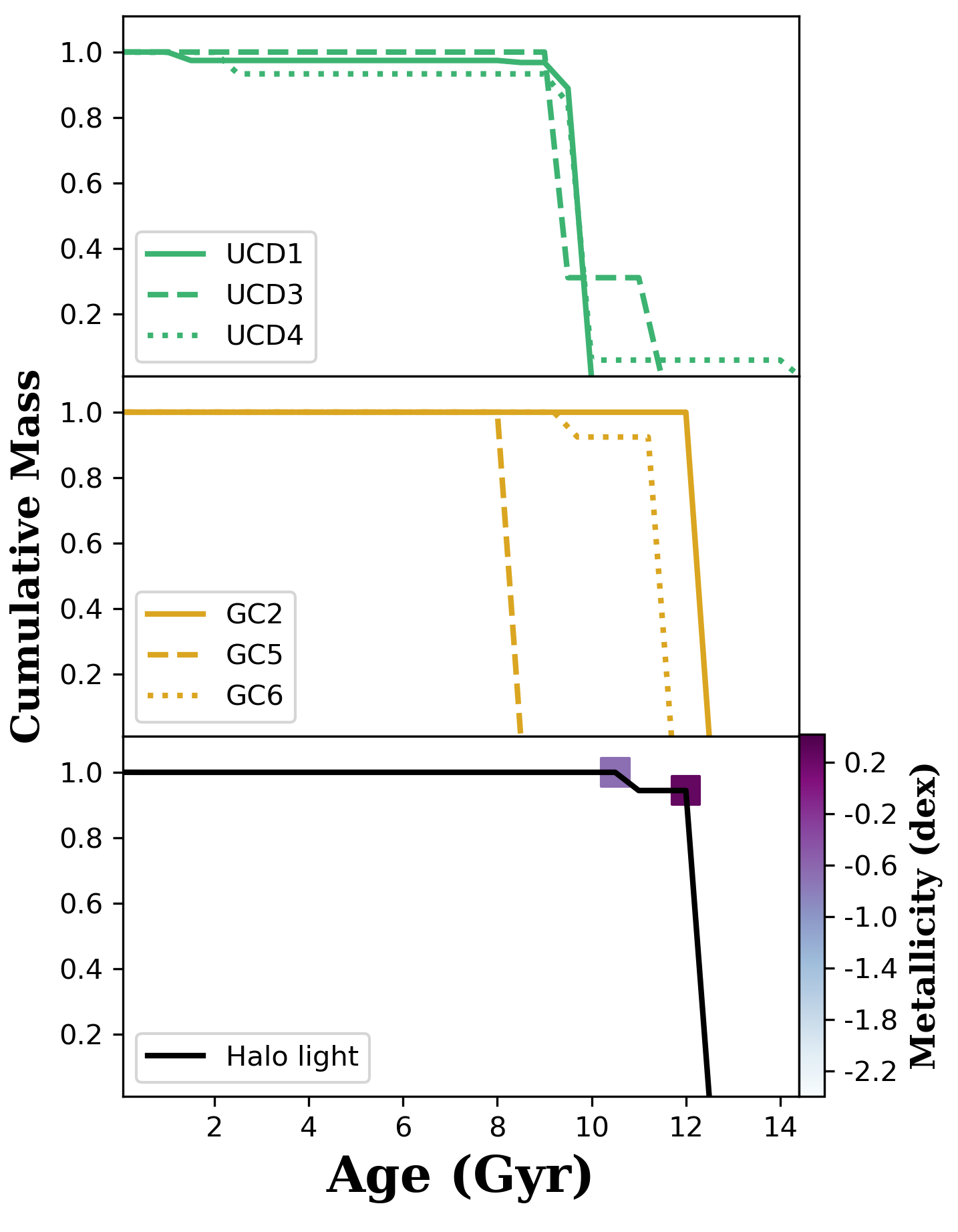}
	\caption{Star formation histories of M87 halo light and compact objects. Lines show the cumulative mass of each as derived from our stellar population analysis. These indicate old stellar populations for GC2 and GC6, and more extended star formation histories with younger mean ages for the UCDs and GC5. 
		For the halo light, filled squares coloured-coded by metallicity show the dominant $\sim$80\% mass fraction (age = 12.0 Gyr) and the secondary $\sim$20\% mass fraction (age = 10.5 Gyr). 
	}
	\label{fig:sfh}
\end{figure}

\subsection{Results}


Fig.~\ref{fig:sfh} shows the derived star formation histories (SFHs; i.e. the build-up of stellar mass over cosmic time) 
for the halo light and the compact objects for which a stellar population analysis was possible (objects 1 to 6). 
The halo light has 
super-solar alpha-element enhancement 
and evidence of two components, i.e 
a dominant component of about 80\% by mass and a secondary of about 20\% by mass from our full-spectral fitting analysis. The dominant halo stellar population is 12.0 Gyr old and metal-rich ([Z/H] = +0.16\,dex). Whereas the secondary component is also old (10.5 Gyr) but more metal-poor ([Z/H] = --0.66\,dex). 
The plot shows that GC2 and GC6 are also dominated by old stellar populations, whereas 
the UCDs and GC5 have more extended star formation histories with mass-weighted ages 8--9\,Gyr. A couple of caveats to the above should be mentioned, i.e. the presence of blue horizontal branch stars in low metallicity systems can lead to lower inferred ages 
(\citealt{2000ApJ...541..126M}; \citealt{2018ApJ...854..139C}) and the stellar population parameters of GC6 are highly uncertain due to the low S/N of its spectrum.

In Table 3 we summarise our results listing the measured radial velocities and the derived stellar population parameters.  The stellar population parameters correspond to the mass-weighted mean values, and in the case of the halo light we include both the dominant ($\sim$80\%) and secondary ($\sim$20\%) stellar populations. 
Although our radial velocity measurement for the inner halo of M87 differs somewhat from the value currently quoted in NED (i.e. 1246 $\pm$ 3 km/s),  it is well within the range of velocities reported over the years as listed by Hyperleda (http://leda.univ-lyon1.fr/).  We measure a velocity dispersion at 
56 arcsec (or 4.5 kpc) from the galaxy centre
of $\sigma$ = 252 $\pm$ 4 km/s.


Based on our derived stellar populations (which assume a Kroupa IMF), we have estimated the mass-to-light ratios for objects 1--6 and converted their absolute magnitudes (from Table 2) into stellar masses. The uncertainties quoted on the derived stellar masses are from the stellar population modelling only (the uncertainty from the total luminosities has not been propagated and may contribute another 10-20\%). We find that the stellar masses for the GCs are all greater than 10$^6$ M$_{\odot}$ suggesting they are 
all at the massive tail end of the GC distribution. 

\begin{figure*}
	\centering
	\includegraphics[width=0.9\linewidth]{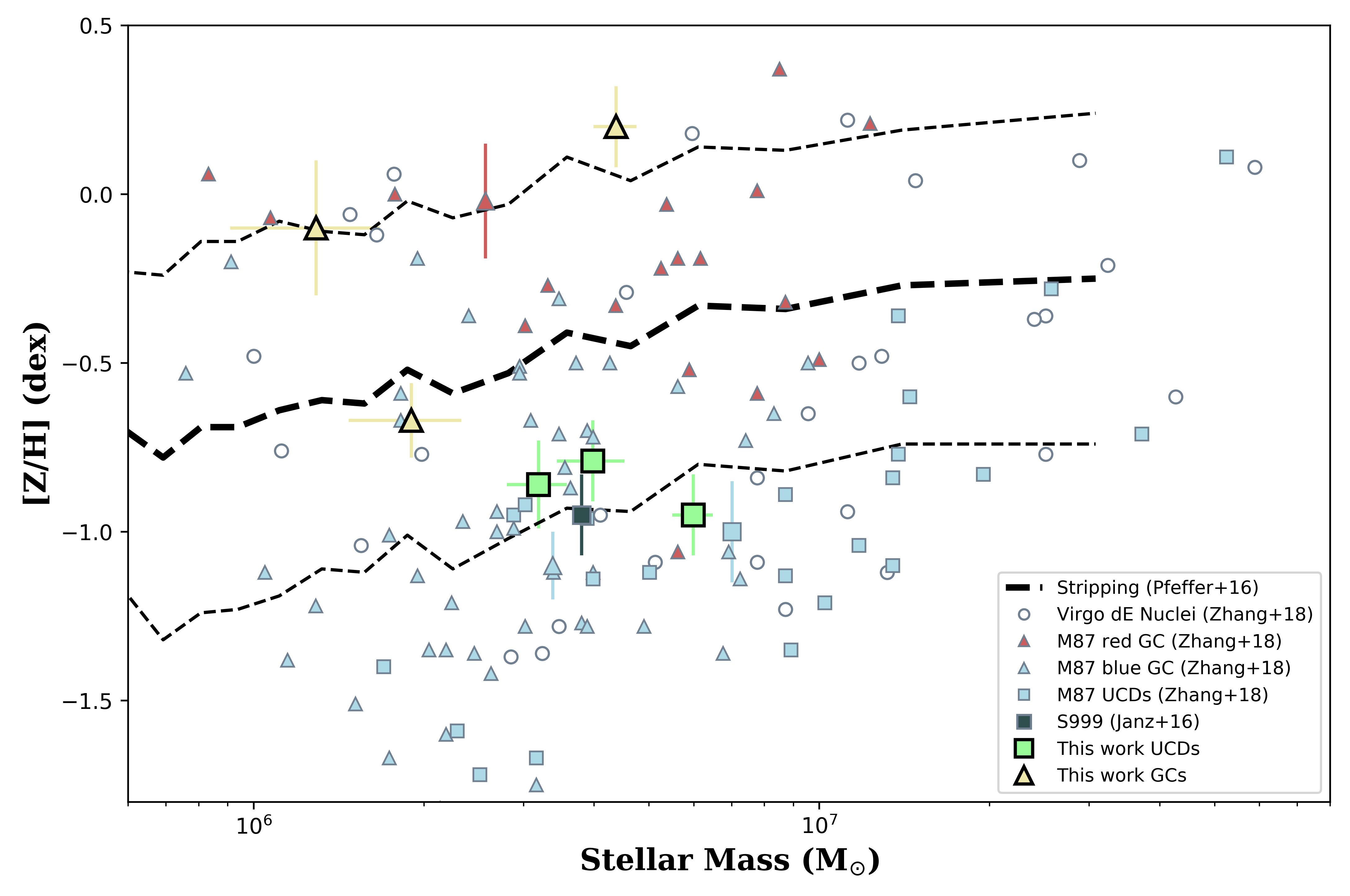}
	\caption{The mass-metallicity relation for a range of objects. 
	Literature samples of dE nuclei, UCDs and GCs are shown, along with the 3 UCDs and 3 GCs from this study. Individual M87 UCDs and GCs from \citet{2018ApJ...858...37Z} are shown along with mean values (larger symbols with error bars). 
	The well studied M87 UCD S999 from 
	\citet{2015MNRAS.449.1716J} is also shown. 
	Our M87 UCDs are located in a similar region of the plot to the literature UCDs. 
	The dashed lines show the galaxy stripping model of \citet{2016MNRAS.458.2492P} for the origin of UCDs. The stripping model overpredicts the metallicity of 
	observed UCDs. 
	}
	\label{fig:mz}
\end{figure*}


The spectra of the two red GCs (2 and 6) in Fig.~\ref{fig:spectra} are similar in appearance to the M87 halo light. 
Indeed, not only do these two red GCs have similar SFHs to the halo light (Fig.~\ref{fig:sfh}), but they also reveal similar metallicities, ages and alpha-element abundances to the dominant stellar population of the halo. With the caveat of projection effects, this suggests that these two red GCs were formed contemporaneously from the same gas as the majority of field stars at this halo location, further strengthening the connection between red GCs and host galaxy field stars seen in their mean colours \citep{2001MNRAS.322..257F}
and their kinematics 
\citep{2013MNRAS.428..389P}. 

All three UCDs, and the blue GC5 have old ages ($\sim$8--9 Gyr) and low metallicities (--0.7 $>$ [Z/H] $>$ --1.0) with solar alpha-element abundances (with the exception of  
UCD1, with ([$\alpha$/Fe] = +0.40). Thus they have some properties that are similar to the 20\% mass component of the halo light at the same (projected) radius.

\begin{figure*}
	\centering
	\includegraphics[width=0.9\linewidth]{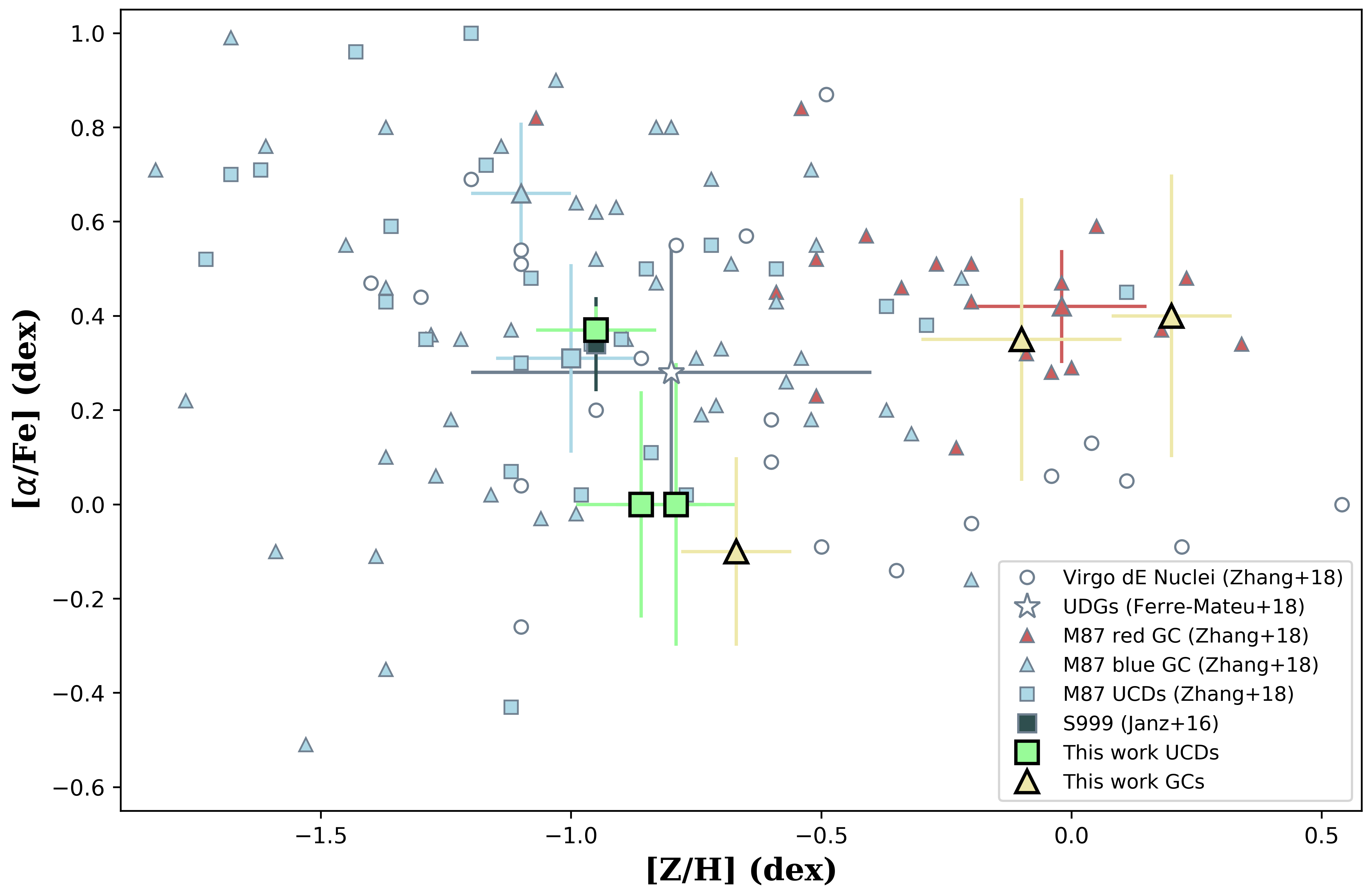}
	\caption{The alpha-element metallicity relation for a range of objects. 
	The plot shows Virgo dE nuclei, UCDs and GCs \citep{2018ApJ...858...37Z} and Coma cluster ultra diffuse galaxies \citep{2018MNRAS.479.4891F}, 
	along with the 3 UCDs and 3 GCs from this study. The well studied M87 UCD S999 from 
	\citet{2015MNRAS.449.1716J} is also shown. The M87 GCs and UCDs from this work lie within the range of values for literature GCs and UCDs, and those of dE nuclei and the ultra diffuse galaxies. 
	}
	\label{fig:alpha}
\end{figure*}

\section{Discussion}

UCDs are generally thought to be 
remnant nucleus of a dwarf galaxy that has undergone tidal stripping near a 
more massive galaxy \cite[see discussion in][]{Brodie2011,Forbes2014,Norris2015}. 
This stripping of a nucleated galaxy, which leaves only the nucleus (to be classified as a UCD) has been modelled by \cite{2003MNRAS.344..399B}, \citet{Pfeffer2013} and \cite{2016MNRAS.458.2492P}. 
Based on elevated mass-to-light (M/L) ratios \citep{Forbes2014}, extended star formation histories \citep{Norris2014}, kinematics \citep{Norris2011}, the detection of massive black holes \citep{Seth2014} and the presence of  tidal tails \citep{Jennings2015}, the evidence that at least some UCDs have a tidally-stripped galaxy origin is secure.

Other origins for some low mass UCDs are possible
(see discussion in \citealt{2019MNRAS.488.5400N})
For example, they may be simply the massive tail of the GC distribution (perhaps formed via merging of individual GCs or from an extremely massive giant molecular cloud). So while 
the relative contributions of UCDs that are massive GCs versus stripped nuclei 
is still subject to debate, the massive GC origin can not easily explain the large sizes, elevated M/L ratios nor the presence of massive black holes as observed.

Previous studies of the M87 halo have included large samples of UCDs and GCs. For example, \citet{1998ApJ...496..808C} studied 150 UCDs and GCs 
around M87, finding uniformly old ages ($>$ 8 Gyr)  with a large metallicity range from [Fe/H] = --2 to solar. \citet{2010ApJ...724L..64P} obtained the stellar population parameters for 10 UCDs in the Virgo cluster (along with 34 dEs). Their UCDs are generally older than 8 Gyr with a mean age of 11.7 Gyr (which is similar to that for their dEs located in high density regions). Their UCDs have a mean metallicity [Z/H] $\sim$ --1 and a large range in alpha-element ratios from slightly sub-solar to [$\alpha$/Fe] = +0.5. From their dE sample, \citet{2010ApJ...724L..64P} noticed significant age differences in the stellar populations of dEs located in high and low density regions. 

\citet{2018ApJ...858...37Z} combined a Lick line index analysis of their new data with that from the literature including \citet{1998ApJ...496..808C} and \citet{2010ApJ...724L..64P} mentioned above, along with 
\citet{2007MNRAS.378.1036E},  \citet{2007MNRAS.382.1342F},
\citet{2012MNRAS.425..325F} and 
\citet{2016MNRAS.456..617J}. 
This gave them a total sample of 40 UCDs (defined to have R$_e$ $>$ 10 pc) and 118 GCs around M87 with spectra. Using stellar the population models of \citet{2003MNRAS.339..897T} they found a mean alpha-element ratio of [$\alpha$/Fe] = +0.4, ages generally older than 8 Gyr and 
metallicities spanning the range [Z/H] $\sim$ --1.7 to +0.3 for the UCDs. They summarise the mean stellar population properties for their UCDs, red GCs and blue GCs in their table 1. 

In our work we used 
the models of \citet{2016MNRAS.463.3409V}, in contrast to the models of \citet{2003MNRAS.339..897T} used by \citet{2018ApJ...858...37Z}. The 
derived metallicities from these two models can be directly compared as systematic differences between them only begin to occur for [Z/H] $<$ --1.5 
\citep{2007MNRAS.379.1618M}.



In Fig.~\ref{fig:mz} we summarise the stellar mass--metallicity relation for our M87 UCDs and GCs compared to UCDs and GCs from the combined study of \citet{2018ApJ...858...37Z}. 
We also highlight the detailed study by \citet{2015MNRAS.449.1716J}
of the M87 inner halo UCD S999 for which they measured an age of 7.6 Gyr, [Z/H] = --0.95, [$\alpha$/Fe] = +0.34.
They inferred a stellar mass of 3.9 $\times$ 10$^6$ M$_{\odot}$ and a dynamical mass-to-light ratio of 8.2. 
From the figure it is seen that the 3 UCDs studied in our work lie within the range of masses and metallicities inferred for other UCDs around M87. 


We also show in Fig.~\ref{fig:mz} the predicted metallicity range for UCDs formed by galaxy stripping in the model of  \citet{2016MNRAS.458.2492P} after converting their [Fe/H] to [Z/H] assuming an alpha-element ratio of [$\alpha$/Fe] = +0.3. Our UCDs, and those in the literature,  lie within, or just below, the predicted model range. This is also the case for the metallicity range observed for the Virgo dwarf galaxy nuclei (potential progenitors of UCDs). Thus the stripping models of \citet{2016MNRAS.458.2492P} currently appear to over-predict the metallicities of UCDs by about 0.5\,dex. 

 
\begin{figure*}
	\centering
	\includegraphics[width=0.75\linewidth, angle=-90]{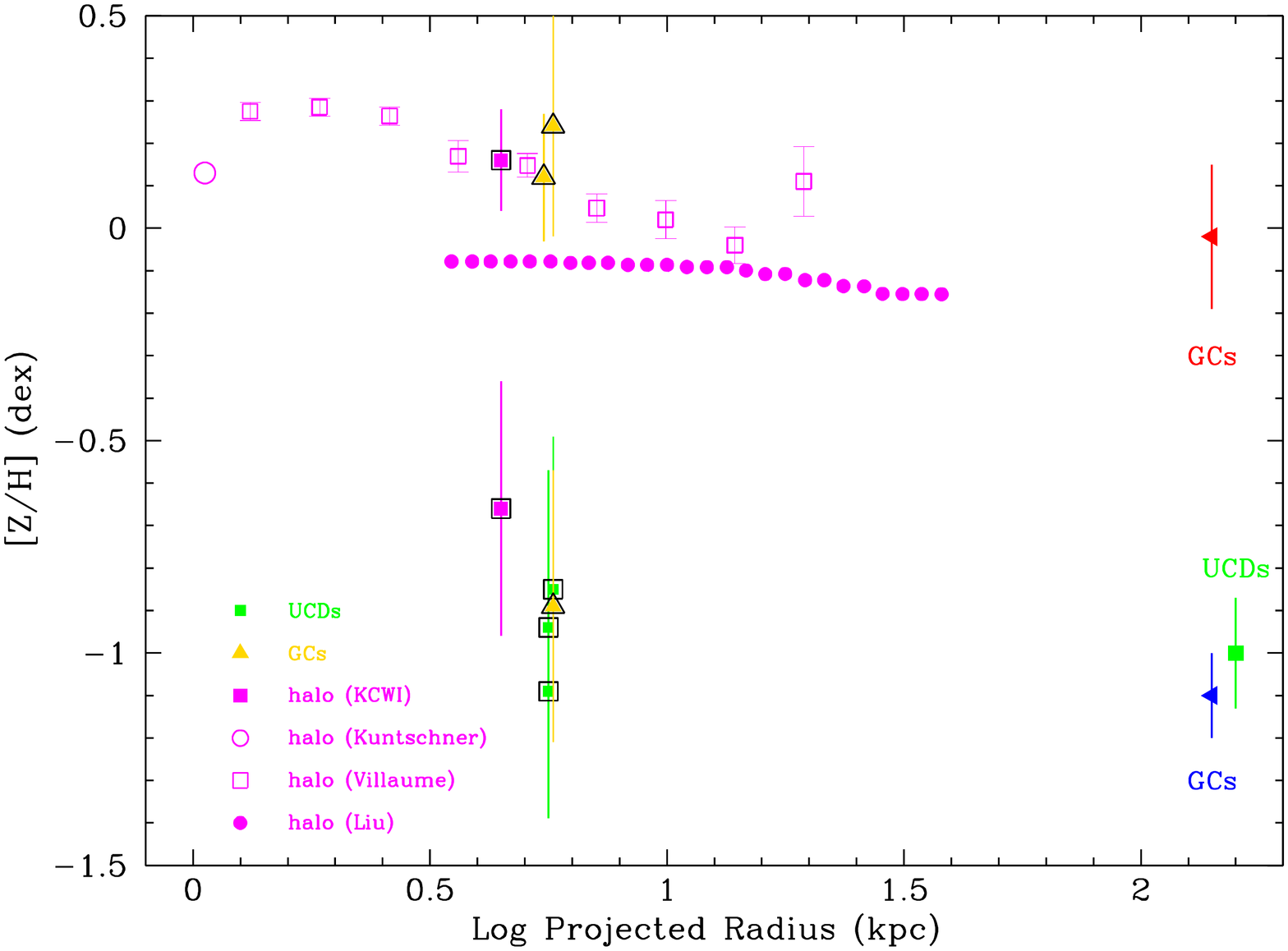}
	\caption{Metallicities in the halo of M87. Galaxy halo stellar metallicities are shown by magenta symbols. Data 
	from Villaume et al. (2019, priv. comm.) are shown as magenta open squares, photometric metallicities from \citet{2005AJ....129.2628L} by magenta filled circles and the single data point from \citet{2010MNRAS.408...97K} by an open circle. The two halo components (representing $\sim$80\% and $\sim$20\% in mass and [Z/H] = +0.16 and --0.66\,dex respectively) found in this work are shown  
	as a magenta filled squares at 4.5 kpc.
	Our KCWI GCs are shown as gold filled triangles and our UCDs as green filled  squares. The mean values of the red and blue luminous GCs and UCDs from \citet{2018ApJ...858...37Z} are shown on the right hand side of the plot at large radii.
	The metal-rich GCs, from our KCWI data, have similar metallicities to the stellar halo starlight at the same projected radius.
	}
	\label{fig:fe}
\end{figure*}

In Fig.~\ref{fig:alpha} we show [$\alpha$/Fe] versus total metallicity. The M87 UCDs and GCs, from our work and from \citet{2018ApJ...858...37Z}, 
tend to have super-solar alpha ratios indicative of rapid star formation. These ratios also cover a similar range to those of Virgo dE nuclei (although  the dE nuclei have a lower mean value). So, similarly to Fig.~\ref{fig:mz}, the M87 UCDs appear to be generally consistent with being the stripped nuclear remnant of a dwarf galaxy. 

As well as nucleated dE galaxies, the progenitor galaxy for a UCD is potentially a nucleated ultra diffuse galaxy, as first suggested by \citet{2017ApJ...839L..17J}. 
A large fraction of ultra diffuse galaxies have nuclei and the space density of such galaxies declines towards the centre of several clusters
(e.g. Alabi et al. 2020, submitted). 
This can be contrasted with the rise in space density of UCDs towards the centre of those clusters,  suggesting one type of galaxy is transformed into another (\citealt{2017ApJ...839L..17J}; \citealt{2019ApJ...887...92J}). Alpha-element abundances are not generally available for Virgo cluster ultra diffuse galaxies but they do exist for Coma cluster ones (see \citealt{2018MNRAS.479.4891F} 
and references therein). We include their mean value in Fig.~\ref{fig:alpha}, and find that Coma cluster ultra diffuse galaxies have [$\alpha$/Fe] ratios that are fully consistent with our M87 UCDs. Thus the progenitor galaxies of UCDs could be either traditional dwarf galaxies or ultra diffuse galaxies with nuclei. 

Further evidence that our 3 UCDs are the remnants of stripped galaxies comes from their physical sizes, which are all over 30 pc. Size was also used recently by 
 \citet{2019A&A...625A..50F} to support the stripped nucleus origin for a UCD they observed in the Fornax cluster. Using MUSE on the VLT, they measured an old age ($>$ 8 Gyr) and low metallicity ([Z/H] = --1.12). 
 Although such values could be consistent with a massive star cluster, they also measured a size from HST imaging of R$_e$ = 24 pc placing this object beyond the regime of GCs,  which typically have sizes of 2--3 pc. 
 
The orbits of UCDs offer yet another diagnostic as to their origin.  \citet{2015ApJ...802...30Z} found that the outer ($>$ 40 kpc) M87 UCDs tend to be on radially biased orbits, which are conducive to tidal stripping. In the inner regions the UCDs were predominately on tangentially-biased orbits, which may indicate that those on radial orbits have been disrupted. Detailed modelling is required to draw firm conclusions. 

\begin{figure*}
	\centering
	\includegraphics[width=0.75\linewidth, angle=-90]{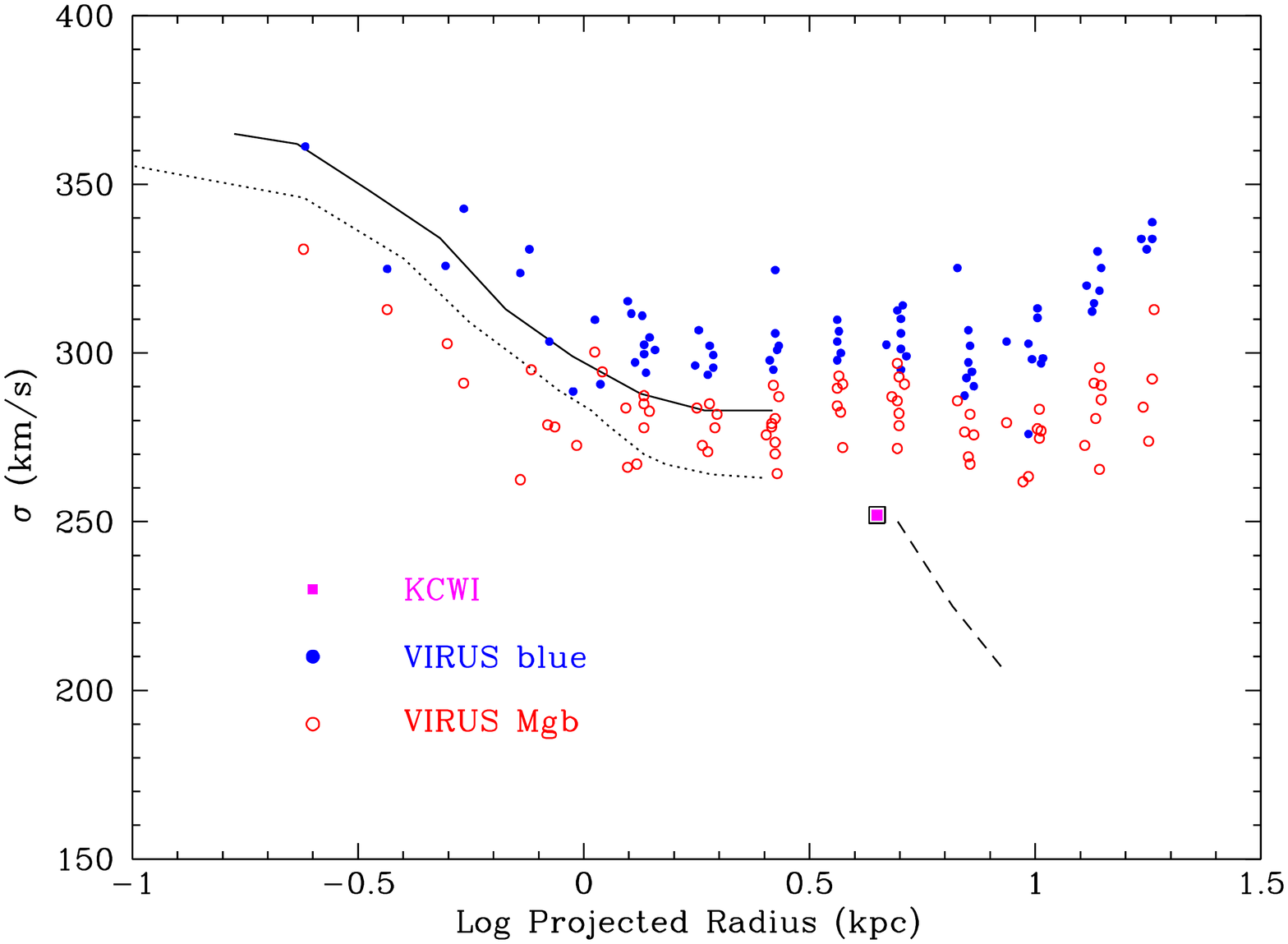}
	\caption{Radial velocity dispersion profile. Our measurement of the velocity dispersion at 4.5 kpc from KCWI (magenta square) is compared to values from the literature. The inner solid line is from the SAURON project \citep{2004MNRAS.352..721E}, the inner dotted line is from \citet{2014MNRAS.445L..79E} using the MUSE instrument and the outer dashed line from the SLUGGS project \citep{2016MNRAS.457..147F}. The blue filled circles show data from the VIRUS instrument when velocity dispersion is measured from blue absorption lines, and the red open circles when it is measured from the Mg$_b$ spectra region at $\sim$5200\AA~\citep{2011ApJ...729..129M}. Our KCWI data support a declining velocity dispersion profile at large radii.  }

	\label{fig:sigma}	
\end{figure*}	

Similar to other large galaxies, M87 reveals colour (and hence metallicity) radial gradients in both GC subpopulations and its field stars \citep{2018MNRAS.479.4760F}. 
In Fig.~\ref{fig:fe}, we compare our measurements to those from the literature for the metallicity of M87's halo stars over a range of galactocentric radius. 
As part of the SAURON project \citep{2010MNRAS.408...97K} studied the stellar population of M87 including its metallicity. They quote a value of [Z/H] = +0.13 $\pm$ 0.02\,dex at R$_e$/8 (which corresponds to 13.3 arcsec). They also quote a metallicity value at R$_e$, however this is an extrapolation from their last measured value to 106 arcsec (i.e. well beyond the radial limit of their data) and we do not use it here for that reason. We include 
the unpublished profile 
of Villaume et al. (2020, priv. comm.) after converting their [Fe/H] into [Z/H] assuming a constant 0.3 dex additive term. We also include the profile from \citet{2005AJ....129.2628L} which is based on their conversion of their 13 band photometry into [Z/H] metallicity. The systematics associated with the Liu et al. approach compared to traditional spectroscopic metallicities has not been quantified in the literature.  
Our inner halo metallicity measurement for the dominant (80\% by mass) component is in good agreement with the halo metallicity profile of 
Villaume et al. at the same radius (4.5 kpc). 
We also show the 20\% by mass halo component, GCs and UCDs in Fig.~\ref{fig:fe}.  The two red,  metal-rich GCs are consistent with the halo stellar metallicity profile. The one blue, metal-poor GC has a much lower metallicity, similar to those of our 3 UCDs.

We also show on Fig.~\ref{fig:fe} the mean metallicity of luminous red and blue GCs from \citet{2018ApJ...858...37Z}, which we plot as upper limits in radius at their quoted enclosed maximum radius of 140 kpc ($\sim$20R$_e$), although half of them are located within 50 kpc. The mean value for the \citet{2018ApJ...858...37Z} UCDs is shown at an arbitrary large radius. 

The starlight halo metallicity profile of Villaume et al. (2020, priv. comm.) shows a negative gradient beyond 3 kpc but a hint of a rise again at the outermost data point. Whereas the profile of \citet{2005AJ....129.2628L} is shallower with no sign of an upturn at large radii. 
Observations deep into the halo of M87 beyond 30 kpc (i.e. $>$5R$_e$) are needed to probe whether the halo metallicity remains relatively metal-rich 
(as seen in the Sombrero galaxy;  \citealt{2020arXiv200101670C}) 
or whether it approaches much lower metallicities as seen in some other nearby galaxies at large radii (e.g. NGC~3115; \citealt{Peacock2015}).


The velocity dispersion profile in the halo of M87 has been the subject of debate in the literature, i.e. whether it is falling, flat or rising with radius. The absolute value of the velocity dispersion and the shape of its profile have strong implications for any derived mass profile and inferred orbital anisotropy. The MASSIVE survey of massive early-type galaxies \citep{2018MNRAS.473.5446V} found that, for radii greater than 5 kpc, velocity dispersion profiles were either falling, flat or rising in equal proportions in their sample (which did not include M87). As part of our stellar population analysis we measure the velocity dispersion of the M87 halo light. We now compare this one data point to the kinematic profile of M87 from the literature. 

The radial velocity dispersion profile of M87, obtained by \citet{2004MNRAS.352..721E} as part of the SAURON project, declines out to a radius of $\sim$30 arcsec, with a hint of flattening at their largest radii probed. A similar profile, but shifted to lower velocities, is seen using the MUSE instrument \citep{2014MNRAS.445L..79E}. 
Using the VIRUS instrument, \citet{2011ApJ...729..129M} found a flat profile at large radii when using the Mg$_b$ lines to derive the velocity dispersion but a rising one when using bluer absorption lines. The source of the discrepancy is unknown. Furthermore, using spectra from the SLUGGS project 
\citep{2014ApJ...796...52B}, \citet{2016MNRAS.457..147F} measured the velocity dispersion between 60 and 110 arcsec from the Calcium Triplet lines and found a declining radial profile and an offset of some 50 km/s from the \citet{2011ApJ...729..129M} results. 
We investigate these discrepancies using 
our KCWI measurement of $\sigma$ = 252 $\pm$ 4 km/s at a radius of 54 arcsec.
In Fig.~\ref{fig:sigma} we show the velocity dispersion profiles from the literature along with our independent measurement from KCWI. Our value is consistent with the profile of \citet{2016MNRAS.457..147F} and lies below both sets of values from \citet{2011ApJ...729..129M}. 
Thus, it favours a declining stellar velocity dispersion profile at large radii in M87. 

\section{Conclusions}

Here we present the photometric and stellar population properties for three GCs and three UCDs associated with M87, along with those of M87's inner halo field stars (at a similar projected galactocentric radius of $\sim$5 kpc). For the first time data have been obtained from the same observations (i.e. the integral field unit KCWI on the Keck II telescope) with results obtained from the same stellar population models (thus avoiding any offsets due to different model calibrations). \\

\noindent
We find the following results:\\

\noindent
$\bullet$ The stellar populations in the inner halo of M87 reveal evidence for a dual nature. The dominant population, with a mass-weighted contribution of $\sim$80\%, is old  ($\sim$12\,Gyr) and metal-rich ([Z/H]$\sim +$0.2).
The secondary population, with a mass-weighted contribution of $\sim$20\%, is similarly old 
($\sim$11\,Gyr) but metal-poor ([Z/H]$\sim$--0.7). The M87 halo is  
enhanced in alpha elements.\\

\noindent
$\bullet$ The two red GCs (GC2 and GC6) are old ($\sim$11 Gyr), metal-rich ([Z/H] $\sim$ +0.1) and alpha enhanced ([$\alpha$/Fe] $\sim$ +0.4). This close agreement in age, metallicity and alpha elements with the dominant component of M87's field stars reinforces the  association between halo stars and metal-rich GCs, e.g. they may have formed in-situ from the same initial gas at a similar redshift.\\

\noindent
$\bullet$ The three UCDs, and the one blue GC5,  have similar stellar populations, revealing more extended star formation histories (with mean ages 8--9 Gyr), low metallicities 
(with [Z/H] $\sim$ --0.8) and near solar alpha-elements. \\

\noindent
$\bullet$ The UCDs have similar metallicities, alpha-element ratios and stellar masses to other UCDs around M87. These properties, and physical sizes greater than 30 pc, are consistent with the scenario that they are the stripped remnants of nucleated galaxies (either traditional dwarf galaxies or ultra diffuse galaxies). We also find that the stripping model of \citet{2016MNRAS.458.2492P} over-predicts literature UCD metallicities by $\sim$0.5 dex. \\


\noindent
$\bullet$
The stellar velocity dispersion profile of M87 reveals a large discrepancy between different studies in the literature. This has strong implications for any calculation of its enclosed mass profile. We help to resolve this discrepancy by measuring a velocity dispersion in the inner halo of 252 km/s, which favours the reported declining profile.


\section*{Acknowledgements}

We thank I. Martin-Navarro and A Wasserman for help with the observations, A. Villaume for supplying data ahead of publication and A. Alabi for useful comments. Referee comments have helped to improve the final manuscript. DAF thanks the ARC for financial support via 
DP160101608. 
AFM has received financial support through the Postdoctoral Junior Leader Fellowship Programme from La Caixa Banking Foundation (LCF/BQ/LI18/11630007). AJR was supported by National Science Foundation grant AST-1616710 and as a Research Corporation for Science Advancement Cottrell Scholar.

The data presented herein were obtained at the W. M. Keck Observatory, which is operated as a scientific partnership among the California Institute of Technology, the University of California and the National Aeronautics and Space Administration. The Observatory was made possible by the generous financial support of the W. M. Keck Foundation. The authors wish to recognize and acknowledge the very significant cultural role and reverence that the summit of Maunakea has always had within the indigenous Hawaiian community.  We are most fortunate to have the opportunity to conduct observations from this mountain. 

\section*{Data Availability}

Data are available from the Keck Observatory Archive (KOA): https://www2.keck.hawaii.edu/koa/public/koa.php




\bibliographystyle{mnras}
\bibliography{M87-UCDs} 



\appendix


\bsp	
\label{lastpage}
\end{document}

%% file: extradefs.tex
\makeatletter
\newcommand*{\rom}[1]{\expandafter\@slowromancap\romannumeral #1@}
\makeatother
\makeatletter
\def\blfootnote{\xdef\@thefnmark{}\@footnotetext}
\makeatother
\newcommand{\ssc}{S\'{e}rsic}
